\documentclass[prd,fleqn,twoside,twocolumn,floatfix]{revtex4}%aps,epsfig]{revtex}

\def\be{\begin{eqnarray}}
\def\ee{\end{eqnarray}}
\def\slash#1{#1\!\!\!/}
\usepackage{latexsym,exscale}
\usepackage{epsfig}
\usepackage{amsmath,amssymb,amsfonts}

\begin{document}

\title{Breaking rotational symmetry in two-flavor color
superconductors}
\author{H. M\"uther and A. Sedrakian }-
\affiliation{Institut f\"ur Theoretische Physik, Universit\"at
T\"ubingen, D-72076 T\"ubingen, Germany}

\begin{abstract}
The color superconductivity under flavor asymmetric
conditions relevant to the compact star phenomenology is studied 
within the Nambu-Jona-Lasinio model. We focus on the effect of the deformation 
of the Fermi surfaces on the pairing properties and the energy 
budget of the superconducting state. We find that at finite 
flavor asymmetries  the color superconducting BCS state is 
unstable towards spontaneous  quadrupole deformation of the 
Fermi surfaces of the $d$ and $u$  quarks into ellipsoidal form. 
The ground state of the phase with deformed Fermi surfaces corresponds
to a superposition of prolate and oblate deformed Fermi ellipsoids 
of $d$ and $u$ quarks. 
\end{abstract}

\pacs{12.38-t,26.60.+c,74.20.-z,97.60.Jd,03.75.Fi}

\maketitle
 
\section{Introduction}

A possible outcome of the de-confinement phase transition in 
hadronic matter, when it is compressed in the centers of compact 
stars to densities several times the nuclear saturation density, 
is the formation of a two-flavor  ($u$ and $d$) quark matter
coexisting with leptons under $\beta$ stability and global charge 
neutrality  \cite{GLENDENNING}. It is believed that 
the strong interaction among quarks in the attractive channel(s) pairs 
the $u$ and $d$ quarks of two color (leaving one color unpaired)
to form a  Bardeen-Cooper-Schrieffer (BCS) 
type color superconductor (the so called 2SC 
phase)  \cite{2SC,Rajagopal:2000wf}. In the asymptotic regime of
high-densities rigorous perturbative results can be obtained 
in the weak coupling regime  \cite{Pisarski:1999tv}; at the 
intermediate densities, relevant for compact stars, one has 
to rely upon effective-field theory models of QCD, such as the 
Nambu-Jona-Lasinio (NJL) model.

In compact stars the color superconductivity is likely to occur at 
finite isospin asymmetry, where the $u$ and $d$ quarks fill
two distinct Fermi spheres. The separation of their Fermi energies
is of the order of the electron chemical potential as required by the charge
neutrality and the stability against weak processes  $d\to u
+e+\bar\nu$ and  $ u+e\to d+\nu$, where $e$, $\nu$ and $\bar\nu$ 
refer to electron, electron neutrino and anti-neutrino. Large
separations of the Fermi energies 
(of the order of pairing gap) and corresponding Fermi momenta 
implies an incoherence of the phase space for the $u$ and 
$d$ quarks, which will eventually destroy 
the 2SC superconductivity. If the occupation of the single particle states
of the non-interacting system is preserved when the pairing
interaction is switched on, i.e. the Fermi spheres are filled with 
quasi-particles in an isotropic and homogenous manner, the phase-space
coherence (at finite separations) can be maintained by pairing 
Cooper pairs with finite center of  mass momentum 
[Larkin-Ovchinnikov-Fulde-Ferrel (LOFF) phase] \cite{LOFF}. This is in line with
the basic prescription of the Landau Fermi liquid theory which states
that in  strongly interacting Fermi liquids
the occupation probabilities of interacting quasi-particles 
remain the same as for the non-interacting system when the interactions
are switched on adiabatically.

The loss of the condensation energy 
due to the separation of the Fermi spheres of the $u$ and $d$ quarks, 
however, might favor a non-Fermi liquid occupation of the 
fermionic states, where the
condensation energy is maintained at the cost of extra kinetic energy.
One possibility is the deformation of the Fermi spheres (DFS), 
to the lowest order, in the ellipsoidal form 
(hereafter - DFS phase)  \cite{DFS}. 
While deviations from the spherically
symmetric form of the Fermi surfaces cost extra kinetic energy, the
gain in the condensation energy due to an increase in the phase-space 
overlap between the states that pair can stabilize the system.
Note that  we assume 
that the pairing interaction is isotropic in space; such a deformation  
spontaneously breaks the rotational $O(3)$ symmetry of the original 
Lagrangian down to $O(2)$.
Both LOFF and DFS superconducting phases break the global space
symmetries.  Another possibility that maintains these symmetries 
is the rearrangement of 
the Fermi surface of the $d$-quarks [interior gap pairing             
(IGP)] \cite{IGS}. The IGP assumes $d$-quark holes located in a strip around 
the Fermi surface of the $u$-quarks are lifted to the excited 
states above the Fermi surface of the $d$ quarks and the pairing 
occurs at the shores of the $u$-quark Fermi sea \cite{IGS}.

A robust feature of the 2SC 
superconductors is the appearance of the
crystalline color superconducting (CCS) state (the analog of LOFF
phase) in a wide range of chemical potential 
asymmetries  \cite{Alford:2000ze,Bowers:2001ip,Leibovich:2001xr,Rajagopal:2001yd,Kundu:2001tt,Bowers:2002xr,Bowers:2002pn,Casalbuoni:2001gt,Casalbuoni:2002hr,Casalbuoni:2002my,Casalbuoni:2002pa}.
In the CCS state the Cooper-pairs carry finite center-of-mass
momentum, their order parameter varies periodically in space and, hence, 
the CCS spontaneously breaks the global space 
symmetries of the original BCS 
state \cite{Alford:2000ze,Bowers:2001ip,Leibovich:2001xr,Rajagopal:2001yd,Kundu:2001tt,Bowers:2002xr,Bowers:2002pn,Casalbuoni:2001gt,Casalbuoni:2002hr,Casalbuoni:2002my,Casalbuoni:2002pa}.
One of the consequences of the broken space symmetries is the
existence of new massless Goldstone modes (phonons) in the CCS phase
 \cite{Casalbuoni:2001gt,Casalbuoni:2002hr,Casalbuoni:2002my,Casalbuoni:2002pa};
another consequence is that, the glitches in compact stars
(rotational anomalies seen in the timing data from  pulsars) 
could originate in the CCS phase, if the compact stars feature such a phase
 \cite{Alford:2000ze}.       

This paper studies numerically  the color superconducting 
DFS phase; it extends the  previous non-relativistic analysis \cite{DFS} to 
relativistic systems and interactions specific to the color
superconductors. The main result of this analysis is that the 
spatially homogenous 2SC superconductor is unstable towards formation 
of the DFS phase with quadrupole deformed Fermi surfaces. Clearly, 
this observation does not identify the true ground state of a 
2SC superconductor. Studies of the interplay between different 
non-BCS phases sketched above are needed to answer this question.
For applications to compact stars these phases need to be 
studied under $\beta$ equilibrium conditions (for example see 
 \cite{Iida:2002ev,Steiner:2002gx,Alford:2002kj}).
    
The paper is organized as follows. In Sec. 2, starting from the NJL
Lagrangian in the chirally symmetric phase, we derive the
thermodynamic potential of the flavor asymmetric 2SC phase at finite 
temperatures. Self-consistent equations for the gap and the partial 
densities of the $u$ and $d$ quarks are obtained. The numerical 
solutions of these equation and the resulting phase diagram 
of the color superconducting DFS phase are shown in Sec. 3.
Sec. 4 contains our conclusions.

\section{Thermodynamics of color superconducting DFS phase}

Our treatment is based upon the   Nambu-Jona-Lasinio (NJL) with 
two flavors $(N_f = 2)$ and three colors $(N_c = 3)$   \cite{ALKOFER}. 
We shall assume from the outset that the color superconducting 
phase emerges in the chirally symmetric phase.  The Lagrangian
density of the model is then given by 
\be\label{eq1}
{\cal L}_{\rm eff} &=& \bar \psi (x)
(i\gamma^{\mu}\partial_{\mu})\psi(x)\nonumber\\
&+&G_1(\psi^TC\gamma_5\tau_2\lambda_A\psi(x))^{\dagger}
(\psi^TC\gamma_5\tau_2\lambda_A\psi(x)),\nonumber\\
\ee
where $C=i\gamma^2\gamma^0$ is the charge conjugation operator,
$\tau_2$ is the second component of the Pauli matrix acting in the
SU$(2)_f$ flavor space, $\lambda_A$ is the antisymmetric Gell-Mann
matrix acting in the  SU$(3)_c$ color space. The coupling constant
$G_1$ stands for the four-fermion contact interaction. 
For the sake of simplicity we neglect the effects of the 
mass gap in the quark-anti-quark channel below the critical line for 
the chiral phase transition on the Cooper pairing. The
generalization of the Lagrangian (\ref{eq1}), which 
leads to coupled gap equations in the di-quark and Cooper channels,
is straight forward.  The common Ansatz for the order parameter 
in the 2SC phase is 
\be\label{eq2} 
\Delta \propto \langle \psi^T(x)C\gamma_5\tau_2\lambda_2\psi(x)\rangle,
\ee
where $\lambda_2$ is the second component of the Gell-Mann matrix. 
The Ansatz for the order parameter [Eq. (\ref{eq2})] implies that the color
SU$(3)_c$ symmetry is reduced to SU$(2)_c$ since only two of the quark 
colors are involved in the pairing while the third color remains
unpaired. More complicated Ansaetze would allow for a spin 1 pairing 
of the quarks of the remaining color \cite{Buballa:2002wy,Alford:2002kj2}.

As is well known, the gap equation and the partial densities of 
the up and down quarks can be found from the fixed points 
of the thermodynamic potential density $\Omega$: 
\be\label{eq3} 
\frac{\partial\Omega}{\partial\Delta} = 0, \quad
-\frac{\partial\Omega}{\partial\mu_f} = \rho_f;
\ee
the flavor index $f=u,d$ refers to up ($u$) and down ($d$) quarks.
In the superconducting state the self-consistent solution of 
equations (\ref{eq3}) corresponds to a minimum of the thermodynamic 
potential of the system. 
For the Lagrangian density defined by Eq. (\ref{eq1}) and the pairing 
channel Ansatz Eq. (\ref{eq2}), the finite temperature thermodynamical 
potential $\Omega$ per unit volume is
\begin{widetext}
\be\label{eq3a}
\Omega(\beta\mu)= -\frac{1}{\beta}\sum_{\omega_n}\int\frac{d^3p}{(2\pi)^3}
\frac{1}{2}{\rm Tr}~{\rm ln}\left[\beta
\left(\begin{array}{cc}
 S_{11}^{-1}(i\omega_n, \vec p) & S_{12}^{-1}(i\omega_n, \vec p) \\
S_{21}^{-1}(i\omega_n, \vec p) & S_{22}^{-1}(i\omega_n, \vec p)
\end{array}\right)
\right]+\frac{\Delta^2}{4G_1},
\ee
where $\beta$ is the inverse temperature. The matrix structure of the 
inverse Matsubara propagators $S_{kl}^{-1}(i\omega_n, \vec p)$, ($l,k
=1,2) $
reflects the Nambu-Gor'kov extension 
of the particle-hole space to account for the pair correlations. 
The elements of the Nambu-Gor'kov matrix 
are $2\times 2$ matrices defined as 
 diag$\,[S_{11}^{-1}]=(\slash p +\mu_u\gamma^0,\slash p
+\mu_d\gamma^0 )$, and  diag$\,[S_{22}^{-1}] 
=(\slash p-\mu_u\gamma^0,\slash p-\mu_d\gamma^0)$; 
 diag$\,[S_{12}^{-1}]=(\Delta\gamma_5\tau_2\lambda_2,
\Delta\gamma_5\tau_2\lambda_2)$
and 
 diag$\,[S_{21}^{-1}]=(-\Delta^*\gamma_5\tau_2\lambda_2,
-\Delta^*\gamma_5\tau_2\lambda_2)$
with off-diagonal elements zero. 
Upon carrying out
the traces in the spin, flavor and color spaces and the fermionic 
Matsubara summation over the frequencies $\omega_n$ one finds
\be\label{eq4}
\Omega= -2 \int\!\!\frac{d^3p}{(2\pi)^3}
\Biggl\{ 2p +\sum_{ij}
\left[\frac{1}{\beta} {\rm log} \left(1+e^{-\beta\xi_{ij}}\right)
+E_{ij}+\frac{2}{\beta} {\rm log}
\left(1+e^{-\beta s_{ij} E_{ij}}\right)
\right]\Biggr\}+\frac{\Delta^2}{4G_1},
\ee
where the indeces $i,j = (+,-)$ sum over the four branches of the 
paired and unpaired quasiparticle spectra defined, respectively, as 
$\xi_{\pm\pm} = {(p\pm\mu)}\pm\delta\mu$
and 
$
E_{\pm\pm} = \sqrt{(p\pm\mu)^2+\vert \Delta\vert^2}\pm\delta\mu,
$
where $\delta\mu = (\mu_u-\mu_d)/2$ and $\mu = (\mu_u+\mu_d)/2$
with $\mu_u$ and $\mu_d$ being the chemical potentials of the up and 
down quarks; $s_{+j} =1 $ and $s_{-j}={\rm sgn}(p-\mu)$. Expressions
analogous to (\ref{eq4}) were derived in refs. 
 \cite{Berges:1998rc,Schwarz:1999dj,Kitazawa:2002bc,Bedaque:1999nu}
for finite/zero temperature and flavor symmetric/asymmetric cases.
The variations of the thermodynamic potential (\ref{eq4}) provide 
the gap equation
\be \label{eq5a}
\Delta &=& 8G_1 \int\frac{d^3p}{(2\pi)^3}\Bigg\{
\frac{\Delta}{E_{+-}+E_{++}}
\left[
{\rm tanh}\left(\frac{\beta E_{++}}{2}\right)
+{\rm tanh}\left(\frac{\beta E_{+-}}{2}\right)\right]\nonumber\\&&\hspace{3cm}
+\frac{\Delta}{E_{-+}+E_{--}}
\left[
{\rm tanh}\left(\frac{\beta E_{-+}}{2}\right)
+{\rm tanh}\left(\frac{\beta E_{--}}{2}\right)\right]
\Bigg\}, %\nonumber\\
\ee
and the partial densities of the up/down quarks
\be \label{eq5b}
\rho_{u/d} &=& 
\int\frac{d^3p}{(2\pi)^3}\Bigg\{2f(\xi_{-\mp})-2f(\xi_{+\pm})
\mp\left[1\pm\frac{\xi_{--}+\xi_{-+}}{E_{--}+E_{-+}}\right]
{\rm tanh}\left(\frac{\beta E_{--}}{2} \right)
\mp\left[1\mp\frac{\xi_{+-}+\xi_{++}}{E_{+-}+E_{++}}\right]
{\rm tanh}\left(\frac{\beta E_{+-}}{2} \right)\nonumber\\
&&
\pm\left[1\mp\frac{\xi_{--}+\xi_{-+}}{E_{--}+E_{-+}}\right]
{\rm tanh}\left(\frac{\beta E_{-+}}{2} \right)
\pm\left[1\pm\frac{\xi_{+-}+\xi_{++}}{E_{+-}+E_{++}}\right]
{\rm tanh}\left(\frac{\beta E_{++}}{2} \right)
\Bigg\},
\ee
\end{widetext}
where $f(\xi_{ij})$ are the Fermi distribution functions
and the upper/lower sign corresponds to the $u/d$-quarks.
The changes in the thermodynamic potential due to the phase transition 
to the superconducting phase at constant $\beta$ and $\mu_f$ are equivalent 
to the changes in the free energy at constant $\beta$ and $\rho_f$
when the potentials are expressed in terms of appropriate
thermodynamical variables.
The free energy $F$ is related to the thermodynamic potential 
$\Omega$ by the relation $F = \Omega +\mu_u\rho_u+\mu_d\rho_d$.
Below, we choose to minimize the free energy of the system at constant
temperature and density of the matter as a function of the flavor 
asymmetry parameter defined as $\alpha \equiv (\rho_d-\rho_u)
/(\rho_d+\rho_u)$. Thus, fixing the net density of the quark matter
and the flavor asymmetry, fixes the chemical potentials $\mu_f$ 
by means of Eqs. (\ref{eq5a}) and (\ref{eq5b}). The chemical potentials 
are isotropic if the Fermi surfaces of the species are spherical. 
To accommodate the possibility of their deformation we expand the 
chemical potentials in Legendre polynomials with respect to an (arbitrary) 
axis of spontaneous symmetry breaking of the spherical symmetry of the 
Fermi surfaces: 
\be\label{eq8}
\mu_f = \sum_{l=0}^{\infty}\mu_{f,l} P_l({\rm Cos}\,\theta).
\ee 
The $l=0$ terms are constants which merely renormalize the chemical 
potentials from their values for the case of vanishing pairing interactions. 
Since the interactions are translationally invariant, the $l=1$ terms can
not contribute, since they manifestly break the translational 
symmetry. (The breaking of translational 
symmetry can be caused by the kinetic energy contribution, as is the case 
for the LOFF/CCS phases.) We shall keep below the leading order $l=2$ terms
which break the rotational symmetry of the system by deforming the
Fermi surfaces into an ellipsoidal form. The corresponding 
expansion coefficients are treated below as variational parameters to 
minimize the free energy of the color superconductor. 
With the leading order in the deformation terms kept, the chemical
potentials of the $u$ and $d$ quarks can be cast in the simple form
\be 
\mu_f =  \bar\mu_f \left[1+(\varepsilon_S\pm\varepsilon_A) ({\rm Cos}\theta)^2 \right],
\ee
where the lower/upper sign corresponds to up/down quarks;
$\varepsilon_{S/A} =(3/2)[\mu_{2d}/\bar\mu_{0d}\pm \mu_{2u}/\bar\mu_{0u}]$
and $\bar\mu_f = \mu_{f0} -\mu_{f2}/2.$
The parameters $\varepsilon_S$ and $\varepsilon_A$ 
describe the deformations of the Fermi surfaces; 
the symmetric part ($\varepsilon_S$)
is the measure of the conformal expansion/contraction of the
Fermi surfaces; the antisymmetric part ($\varepsilon_A$) 
describes the relative deformations of the Fermi spheres of $u$ and 
$d$ quarks. Note that the expansion (\ref{eq8}) need not 
conserve the volume of a Fermi sphere; the isotropic 
expansion parameters are  self-consistently found 
form the normalization to the same matter 
density as in the case for the undeformed state.

\section{The phase diagram}
 
As the NJL model is non-renormalizable, the momentum integrals in the gap 
equation (\ref{eq5a}) need to be regularized by a cut-off; we employ a 
three-dimensional momentum space cut-off $\vert {\bf p}\vert <
\Lambda$. The phenomenological value of 
the coupling constant $G_1$ in the $\langle q q\rangle $  Cooper
channel is related  to the coupling constant in the $\langle q\bar q\rangle $ 
di-quark channel by the relation $G_1 = N_c/(2N_c-2) G$;
the latter coupling constant and the cut-off
are fixed by adjusting the model to the vacuum properties of the 
system  \cite{Berges:1998rc,Schwarz:1999dj,Buballa:2001wh}.
We employ the parameter set 
$G_1 = 3.10861$ GeV$^{-2}$ and $G_1\Lambda^2 = 1.31$  from Ref. 
\cite{Schwarz:1999dj}.
For this parameter set the value of the gap is 40 MeV at the 
baryonic density $\rho_B=0.31$ fm$^{-3}$ which corresponds to the chemical
potential value $\mu = 320$ MeV in the flavor symmetric state at 
$T=0$.
Figure 1 summarizes the main features of
the DFS phase in the physically relevant regime of flavor 
asymmetries $0.1 \le \alpha \le 0.3 $ likely to occur in
the charge-neutral matter under $\beta$ equilibrium.
The color superconducting gap (left panel) and the 
free energy difference between the superconducting state 
and normal state (right panel) are shown as a function of 
deformation parameter $\varepsilon_A$ for several flavor asymmetries
at the baryonic density $\rho_B = 0.31$ fm$^{-3}$ and 
temperature $T=2$ MeV.
The conformal deformation has been constrained to $\varepsilon_S =0$.
The gap is normalized to its value in the 
flavor symmetric and undeformed state $\Delta_{00} \equiv \Delta
(\alpha = 0, \varepsilon_A =0) = 40$ MeV. The free-energy difference
$\delta F = F_S-F_N$ is likewise normalized to its value  in the 
flavor symmetric and undeformed state $\delta F
(\alpha = 0, \varepsilon_A =0) = 37$ MeV fm$^{-3}$.
The $\beta$ equilibrated quark
matter requires an excess of the $d$ over $u$ quarks, therefore the 
range of the flavor asymmetry is restricted to the positive
values. The deformation parameter $\varepsilon_A$ assumes both
positive and negative values. The main features seen in Fig. 1 are: 
(i) for a fixed $\alpha\neq 0$ and $\varepsilon_A >0$, 
the gap is larger in the DFS state than in the ordinary BCS 
state ($\varepsilon_A=0$); (ii) the minimum of the free energy  
corresponds to the DFS state with  $\varepsilon_A\simeq 0.25$, 
and its position weakly depends on the value of $\alpha$.

\vskip 1.cm

\begin{figure}[htb] % fig 1
\begin{center} 
\includegraphics[angle=0,width=\linewidth]{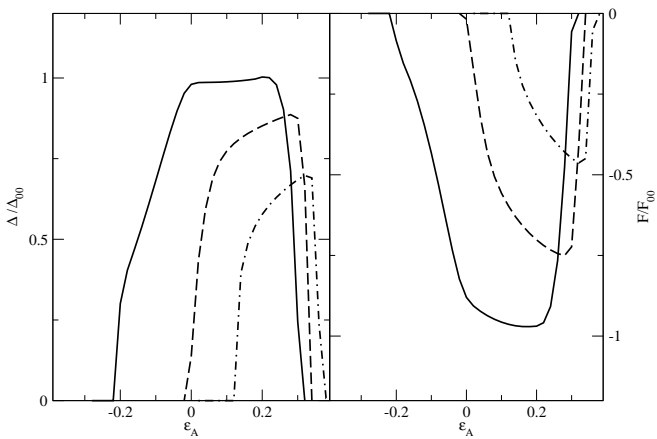}
\end{center}
\caption{
The color superconducting gap (left panel) and the 
free energy (right panel) as a function of 
relative deformation parameter $\varepsilon_A$ for 
the values of the flavor asymmetry $\alpha = 0$ (solid lines), 0.1
(dashed lines),  0.2 (short dashed lines) and 0.3 (dashed dotted
lines) at density $\rho_B = 0.31$ fm$^{-3}$ and temperature $T=2$
MeV.}
\label{MSfig:fig1}
\end{figure}

Positive values of $\varepsilon_A$ imply a prolate
deformation of the Fermi surface of the $d$-quarks and an oblate 
deformation for the Fermi surface of the $u$-quark. 
Figure 2 illustrates the Fermi surfaces for the parameter 
choice $\alpha = 0.3$ and 
$\varepsilon_A = 0.25$, which corresponds to the deformation 
at which the minimum in the free energy is obtained.

\begin{figure}[htb] % fig 2
\begin{center} 
\includegraphics[angle=-0,width=4cm]{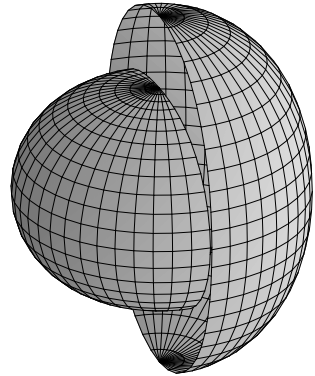}
\end{center}
\caption{
The deformed Fermi surfaces of the $d$ and $u$ quarks for 
the flavor asymmetry $\alpha = 0.3$  and 
deformation parameter $\varepsilon_A = 0.25$. 
The right part of this figure
refers to the Fermi surface of the $d$-quarks, 
exhibiting a prolate deformation,
while the oblate shape on the left corresponds to the Fermi surface of the
$u$-quarks.
}
\label{MSfig:fig2}
\end{figure}

As can be seen from Fig. 2, the deformation of the Fermi surfaces leads
to an area in which the Fermi surfaces of $u$- and $d$-quarks touch each
other, which means that for this part of the phase space there is no blocking
effect reducing the pairing correlation. As a consequence, the superconducting
gap is larger than for the configuration with spherical Fermi surfaces leading
also to more attractive free energy.

The behavior of
the gap as a function of $\alpha$ and $\varepsilon_A$ is best
understood in terms of the symmetric and anti-symmetric combinations 
of the quasiparticle spectra $E_S = (E_{-+}+E_{--})/2$ and 
$E_A = (E_{-+}-E_{--})/2$. Since we assume that $\varepsilon_S=0$
the symmetric combination of the spectra is weakly affected by
the deformation $[\sim {\cal O}(\varepsilon_A\delta\mu)]$.
In the case $E_A = 0$ one recovers the BCS result with perfectly
overlapping Fermi surfaces. The effect of finite $E_A$ at
$\varepsilon_A=0$ is to induce a phase-space de-coherence in the
kernel of the gap equation; this blocking effect reduces the magnitude
of the gap. 

As the chemical potential shift $\delta\mu$ and $\varepsilon_A$
contribute to $E_A$ with different signs, 
switching on finite $\varepsilon_A >0$ acts to reduce $E_A$
and the magnitude of the gap increases due to the restoration of 
the phase-space overlap. 
Once the condition 
$\varepsilon_A > \delta\mu$ is satisfied further increase in the
deformation acts to increase the de-coherence and, thus, decrease  
the magnitude of the gap.

For negative values of $\varepsilon_A$ the 
deformation and the shift in the chemical potentials contribute to
$E_A$ with the same sign, therefore the deformation acts to increase
the phase-space de-coherence which reduces the magnitude of the gap.
These features are seen in Fig. 3, where we show 
the pairing gap as a function of the flavor asymmetry
$\alpha$ and the relative deformation of the Fermi surfaces
$\varepsilon_A$  for the same density and temperature as 
in Fig. 1.   [We assume, as for Fig. 1, that the conformal
expansion/contraction of the Fermi surfaces is 
absent, $\varepsilon_S=0$].

\begin{figure}[htb] % fig 3
\begin{center}            
\includegraphics[angle=-90,width=\linewidth]{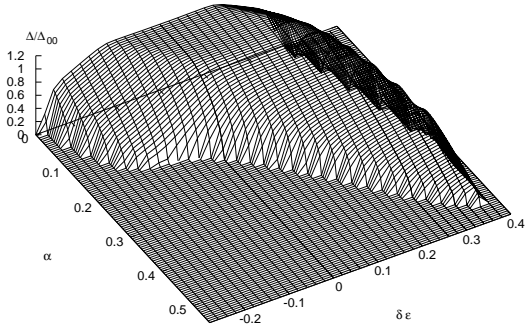}
\end{center}
\caption{The color superconducting gap as a function of 
flavor asymmetry $\alpha$  and the relative deformation parameter 
$\varepsilon_A$ at density $\rho_B = 0.31$ fm$^{-3}$ and temperature $T=2$
MeV. The pairing gap is in units $\Delta (\alpha =0 , \varepsilon_A =
0 ) = 40$ MeV.}
\label{MSfig:fig3}
\end{figure} 
\noindent 

\begin{figure}[htb] % fig 4
\begin{center}
\includegraphics[angle=-90,width=\linewidth]{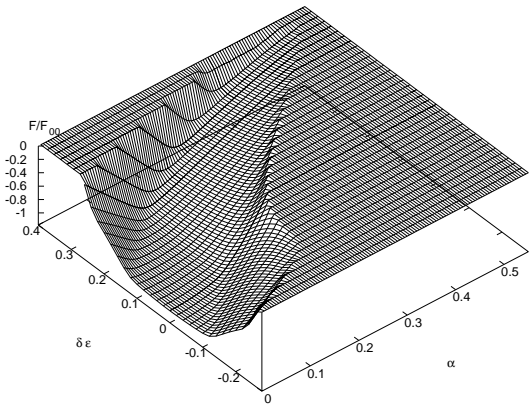}
\end{center}
\caption{The difference between the free energies of the
superconducting and normal phases as a function of 
flavor asymmetry $\alpha$  and the relative deformation parameter 
$\varepsilon_A$. The free energy scale is normalized to its value
$\delta F(\alpha =0, \varepsilon_A=0) = 37$ MeV fm$^{-3}$. 
}
\label{MSfig:fig4}
\end{figure} 
\noindent

Figure 4 shows the difference between the free energies of the
superconducting and normal states $\delta F$ normalized to its value
in the flavor-symmetric BCS state $\delta F_{00} =\delta F(\alpha
= 0, \varepsilon_A = 0)$.  Numerically, the contribution from the 
entropy difference between the normal and superconducting phases is
negligible. Small $\varepsilon_A>0, \alpha\neq 0$ perturbations from 
the flavor symmetric/undeformed state increase $\delta F$
signaling an instability of the BCS state towards deformations of the 
Fermi surfaces. 

For flavor asymmetries $\alpha > 0.3$  and positive values of
the deformation parameter the superconducting state can exists only 
in the DFS phase.
For negative values of $\varepsilon_A$ there is no
energy gain related to the deformation of the Fermi surfaces (except 
a marginal effect in the limit $\alpha \to 0$ and $\varepsilon_A \to
0$). Quite generally the shape of the free energy surface reflects the 
functional behavior of the gap as a function of the parameters
$\alpha$ and $\varepsilon_A$, which is due to the dominant contribution from
the condensation energy (cf. Fig. 1).

\section{Concluding remarks}

Color superconducting 2SC phase becomes unstable towards formation of 
a superconducting phase with deformed Fermi surfaces
of up and down quarks for finite separations of their Fermi levels. 
Small asymmetries in the populations of the of the up/down quarks
are sufficient to achieve the bifurcation point where the form of the 
Fermi surfaces changes from  the spherical to the ellipsoidal form. 
Although the 
departures from the spherical form of the Fermi surfaces costs extra
kinetic energy, there is net energy gain because the deformation 
increases  the phase space coherence between the fermions forming Cooper pairs
and, therefore, the gain in the potential energy.
As mentioned in the introduction, the present study does not establish the
true ground state of the 2SC superconductor under flavor
asymmetry. This can be done by a careful comparison of the
condensation energies of the DFS and CCS phases. With the form of the
lattice of the CCS phase established in the Ginzburg-Landau 
regime \cite{Bowers:2002xr} it becomes feasible to carry out a combined
study of the DFS and CCS phases \footnote{A preliminary computation
with a single plane-wave structure of the CCS phase indicates that a
combination of the CCS and DFS phases is favored.} .

For applications to the compact stars, our treatment 
needs an extention which will include the $\beta$ equilibrium and 
charge neutrality among the quarks and leptons. The onset of the 
DFS phase (or its combinations with non-BCS superconducting
states) will affect the properties of the compact stars in a number 
of ways. One straightforward  implication is the modification of the
specific heat of the superconducting phase which affects the thermal 
cooling of compact stars: we anticipate that the specific heat will 
be suppressed linearly in the DFS phase as compared to the
exponential suppression in the ordinary BCS-state. The spontaneous 
breaking of the rotational symmetry implies emergence of 
massless bosonic modes (Goldstone's theorem). The deformation 
of the Fermi surfaces opens new channels for neutrino radiation via
the bremsstrahlung process of the type $u\to u+\nu+\bar\nu$
and  $d\to d+\nu+\bar\nu$ since the phase-space probability for 
a transition of a quasi-particle from one point on the deformed 
Fermi surface to another  (say, from the equator to the pole)
is non-zero.

%acknowledgments
\section*{}
We thank Michael Buballa and Micaela Oertel for helpful comments 
on the first version of this paper. This work has been supported by the 
Sonderforschungsbereich 382 of DFG.


\begin{thebibliography}{99}
\bibitem{GLENDENNING} N. K. Glendenning, {\it Compact Starts},  
(Springer-Verlag, New York, 1st edition 1996, 2nd edition 2000); 
F. Weber {\it Pulsars as Astrophysical Laboratories for
Nuclear and Particle Physics} (IOP, Bristol 1999); 
{\it Physics of Neutron Star Interiors},  D. Blaschke,
N. K. Glendenning and A. Sedrakian (eds.), (Springer-Verlag, New York, 2001).
\bibitem{2SC}
B. Barrois, Nucl. Phys. {\bf B129}, 390 (1977);
S. Frautschi, Proceedings of workshop on hadronic matter at extreme density,
Erice 1978;
D. Bailin and A. Love, Phys. Rept. {\bf 107}, 325 (1984);
M. Alford, K. Rajagopal and F. Wilczek, Phys. Lett. {\bf B422}, 247 (1998)
[hep-ph/9711395];
R. Rapp, T. Sch\"afer,
E. V. Shuryak and M. Velkovsky, Phys. Rev. Lett. {\bf 81}, 53 (1998)
[hep-ph/9711396].
% \cite{Rajagopal:2000wf}
\bibitem{Rajagopal:2000wf}
For reviews see K.~Rajagopal and F.~Wilczek,
%``The condensed matter physics of QCD,''
in {\it Frontier of Particle Physics / Handbook of QCD}, edited by
M. Shifman (World Scientific Singapore, 2002), chap. 35.
[arXiv:hep-ph/0011333];
M.~G.~Alford,
%``Color superconducting quark matter,''
Ann.\ Rev.\ Nucl.\ Part.\ Sci.\  {\bf 51}, 131 (2001).
%%CITATION = HEP-PH 0011333;%%
% \cite{Pisarski:1999tv}
\bibitem{Pisarski:1999tv}
For example, see R.~D.~Pisarski and D.~H.~Rischke,
%``Color superconductivity in weak coupling,''
Phys.\ Rev.\ D {\bf 61}, 074017 (2000)
[arXiv:nucl-th/9910056] and refrences therein.
%%CITATION = NUCL-TH 9910056;%%
\bibitem{LOFF}
A. I. Larkin and Yu. N. Ovcihnnikov,  Zh. Eksp. Teor. Fiz. {\bf 47}, 
1136 (1964) [Sov. Phys. JETP {\bf 20}, 762 (1965)];
P. Fulde and R. A. Ferrell, Phys. Rev. {\bf 135}, A550 (1964).
\bibitem{DFS} H. M\"uther and A. Sedrakian, Phys.\ Rev.\ Lett. {\bf
88}, 252503 (2002)
[arXiv:cond-mat/0202409];  Phys.\ Rev.\ C.
{\bf 67}, 015802 (2003) [arXiv:nucl-th/0209061].
\bibitem{IGS} W. V. ~Liu and F. Wilczek, arXiv:cond-mat/0208052.
% \cite{Alford:2000ze}   
\bibitem{Alford:2000ze}
M.~G.~Alford, J.~A.~Bowers and K.~Rajagopal,
%``Crystalline color superconductivity,''
Phys.\ Rev.\ D {\bf 63}, 074016 (2001)
[arXiv:hep-ph/0008208].
%%CITATION = HEP-PH 0008208;%%
% \cite{Bowers:2001ip}
\bibitem{Bowers:2001ip}
J.~A.~Bowers, J.~Kundu, K.~Rajagopal and E.~Shuster,
%``A diagrammatic approach to crystalline color superconductivity,''
Phys.\ Rev.\ D {\bf 64}, 014024 (2001)
[arXiv:hep-ph/0101067].
%%CITATION = HEP-PH 0101067;%%
% \cite{Leibovich:2001xr}
\bibitem{Leibovich:2001xr}
A.~K.~Leibovich, K.~Rajagopal and E.~Shuster,
%``Opening the crystalline color superconductivity window,''
Phys.\ Rev.\ D {\bf 64}, 094005 (2001)
[arXiv:hep-ph/0104073].
%%CITATION = HEP-PH 0104073;%%
% \cite{Rajagopal:2001yd}
\bibitem{Rajagopal:2001yd}
K.~Rajagopal,
%``Crystalline color superconductivity,''
AIP Conf.\ Proc.\  {\bf 602}, 339 (2001)
[Nucl.\ Phys.\ A {\bf 702}, 25 (2002)]
[arXiv:hep-ph/0109135].
%%CITATION = HEP-PH 0109135;%%
% \cite{Kundu:2001tt}
\bibitem{Kundu:2001tt}
J.~Kundu and K.~Rajagopal,
%``Mass-induced crystalline color superconductivity,''
Phys.\ Rev.\ D {\bf 65}, 094022 (2002)
[arXiv:hep-ph/0112206].
%%CITATION = HEP-PH 0112206;%%
% \cite{Bowers:2002xr}
\bibitem{Bowers:2002xr}
J.~A.~Bowers and K.~Rajagopal,
%``The crystallography of color superconductivity,''
Phys.\ Rev.\ D {\bf 66}, 065002 (2002)
[arXiv:hep-ph/0204079].
%%CITATION = HEP-PH 0204079;%%
% \cite{Bowers:2002pn}
\bibitem{Bowers:2002pn}
J.~A.~Bowers and K.~Rajagopal,
%``The crystallography of color superconductivity,''
arXiv:hep-ph/0209168.
%%CITATION = HEP-PH 0209168;%%
% \cite{Casalbuoni:2001gt}
\bibitem{Casalbuoni:2001gt}
R.~Casalbuoni, R.~Gatto, M.~Mannarelli and G.~Nardulli,
%``Effective field theory for the crystalline colour superconductive phase  of QCD,''
Phys.\ Lett.\ B {\bf 511}, 218 (2001)
[arXiv:hep-ph/0101326].
%%CITATION = HEP-PH 0101326;%%
% \cite{Casalbuoni:2002hr}
\bibitem{Casalbuoni:2002hr}
R.~Casalbuoni, R.~Gatto and G.~Nardulli,
%``Crystalline color superconductivity: Effective Lagrangian and phonon  dispersion law,''
Phys.\ Lett.\ B {\bf 543}, 139 (2002)
[arXiv:hep-ph/0205219].
%%CITATION = HEP-PH 0205219;%%
% \cite{Casalbuoni:2002my}
\bibitem{Casalbuoni:2002my}
R.~Casalbuoni, E.~Fabiano, R.~Gatto, M.~Mannarelli and G.~Nardulli,
%``Phonons and gluons in the crystalline color superconducting phase of QCD,''
arXiv:hep-ph/0208121.
%%CITATION = HEP-PH 0208121;%%
% \cite{Casalbuoni:2002pa}
\bibitem{Casalbuoni:2002pa}
R.~Casalbuoni, R.~Gatto, M.~Mannarelli and G.~Nardulli,
%``Anisotropy parameters for the effective description of crystalline  color superconductors,''
Phys.\ Rev.\ D {\bf 66}, 014006 (2002)
[arXiv:hep-ph/0201059].
%%CITATION = HEP-PH 0201059;%%
% \cite{Iida:2002ev}
\bibitem{Iida:2002ev}
K.~Iida and G.~Baym,
%``Superfluid phases of quark matter. III: Supercurrents and vortices,''
Phys.\ Rev.\ D {\bf 66}, 014015 (2002).
%[arXiv:hep-ph/0204124].
%%CITATION = HEP-PH 0204124;%%
% \cite{Steiner:2002gx}
\bibitem{Steiner:2002gx}
A.~W.~Steiner, S.~Reddy and M.~Prakash,
%``Color-neutral superconducting quark matter,''
Phys.\ Rev.\ D {\bf 66}, 094007 (2002)
[arXiv:hep-ph/0205201].
%%CITATION = HEP-PH 0204001;%%
% \cite{Alford:2002kj}
\bibitem{Alford:2002kj}
M.~Alford and K.~Rajagopal,
%``Absence of two-flavor color superconductivity in compact stars,''
JHEP {\bf 0206}, 031 (2002)
[arXiv:hep-ph/0204001].
%%CITATION = HEP-PH 0205201;%%
\bibitem{ALKOFER} For a review see R. Alkofer and H. Reinhardt, ``Chiral Quark
Dynamics'', (Springer-Verlag, New York, 1995).
% \cite{Berges:1998rc}
% \cite{Buballa:2002wy}
\bibitem{Buballa:2002wy}
M.~Buballa, J.~Hosek and M.~Oertel,
%``Anisotropic admixture in color-superconducting quark matter,''
arXiv:hep-ph/0204275.
%%CITATION = HEP-PH 0204275;%%
% \cite{Alford:2002kj}
\bibitem{Alford:2002kj2}
M.~G.~Alford, J.~A.~Bowers, J.~M.~Cheyne and G.~A.~Cowan,
%``Single color and single flavor color superconductivity,''
arXiv:hep-ph/0210106.
%%CITATION = HEP-PH 0210106;%%
\bibitem{Berges:1998rc}
J.~Berges and K.~Rajagopal,
%``Color superconductivity and chiral symmetry restoration at nonzero  baryon density and temperature,''
Nucl.\ Phys.\ B {\bf 538}, 215 (1999)
[arXiv:hep-ph/9804233].
%%CITATION = HEP-PH 9804233;%%
% \cite{Schwarz:1999dj}
\bibitem{Schwarz:1999dj}
T.~M.~Schwarz, S.~P.~Klevansky and G.~Papp,
%``The phase diagram and bulk thermodynamical quantities in the NJL model  at finite temperature and density,''
Phys.\ Rev.\ C {\bf 60}, 055205 (1999)
[arXiv:nucl-th/9903048].
%%CITATION = NUCL-TH 9903048;%%
% \cite{Buballa:2001wh}
\bibitem{Buballa:2001wh}
M.~Buballa, J.~Hosek and M.~Oertel,
%``Self-consistent parametrization of the two-flavor isotropic  color-superconducting ground state,''
Phys.\ Rev.\ D {\bf 65}, 014018 (2002)
[arXiv:hep-ph/0105079].
%%CITATION = HEP-PH 0105079;%%
% \cite{Kitazawa:2002bc}
\bibitem{Kitazawa:2002bc}
M.~Kitazawa, T.~Koide, T.~Kunihiro and Y.~Nemoto,
%``Chiral and color superconducting phase transitions with vector  interaction in a simple model,''
arXiv:hep-ph/0207255.
%%CITATION = HEP-PH 0207255;%%
% \cite{Bedaque:1999nu}
\bibitem{Bedaque:1999nu}
P.~F.~Bedaque,
%``Color superconductivity in asymmetric matter,''
Nucl.\ Phys.\ A {\bf 697}, 569 (2002)
[arXiv:hep-ph/9910247].
%%CITATION = HEP-PH 9910247;%%
\end{thebibliography}
\end{document}